# ECONOMIC COMPLEXITY: WHY WE LIKE "COMPLEXITY WEIGHTED DIVERSIFICATION"

Luciano Pietronero[1,2], Andrea Gabrielli[1,2] and Andrea Zaccaria[2]
[1]Physics Department, Sapienza University of Rome (Italy)
[2]Institute for Complex Systems,-CNR, Rome (Italy)

December 23, 2019

Abstract:

A recent paper by Hausmann and collaborators (1) reaches the important conclusion that Complexity-weighted diversification is the essential element to predict country growth. We like this result because Complexity-weighted diversification is precisely the first equation of the Fitness algorithm that we introduced in 2012 (2,3). However, contrary to what is claimed in (1), it is incorrect to say that diversification is contained also in the ECI algorithm (4). We discuss the origin of this misunderstanding and show that the ECI algorithm contains exactly zero diversification. This is actually one of the reasons for the poor performances of ECI which leads to completely unrealistic results, as for instance, the derivation that Qatar or Saudi Arabia are industrially more competitive than China (5,6). Another important element of our new approach is the representation of the economic dynamics of countries as trajectories in the GDPpc-Fitness space (7-10). In some way also this has been rediscovered by Hausmann and collaborators and renamed as "Stream plots", but, given their weaker metrics and methods, they propose it to use it only for a qualitative insight, while ours led to quantitative and successful forecasting. The Fitness approach has paved the way to a robust and testable framework for Economic Complexity resulting in a highly competitive scheme for growth forecasting (7-10). According to a recent report by Bloomberg (9): The new Fitness method, *systematically outperforms standard methods, despite requiring much less data*".

# 1. Introduction: the Grand Challenge of China Economic Growth

Economic growth strongly affects the quality of life of people, but "*growth is devilishly hard to predict*" (The Economist, Jan. 9[th], 2016; see also Ref. 9). The unprecedented economic growth of China in the past 30 years represents an extremely interesting example with great implications and challenges for economic theory and political planning. Is China some sort of inexplicable outlier or can it be cast in some rational conceptual scheme? The debate is hot. Godfree Roberts (https://quora.com/profile/Godfree-Roberts) collected a large number of claims from authoritative sources from 1990 up to present. Each year from 1990 on, all these claims predicted a "hard landing" of China growth. However today the Chinese Gross Domestic Product per capita (GDPpc) is still growing at a rate of about 6.5%. The problem is that China had a growth of more than 5% (sometimes much more) for the past 40 years. This is an absolute anomaly hard to reconcile with the standard economic and statistical criteria, according to which fast growth should not last for more than about ten years (Ref. 11).

These and many other examples show that, in spite of the availability of Big Data and a great effort by many people and institutions, the problem of monitoring and predicting country growth represents a fundamental challenge from both the scientific and economic points of view. In this context China, more than an exception, should represent a fundamental benchmark for any theory or model which attempts to explain and forecast Economic Growth.

One of the early data-driven indicators of industrial competitiveness was the Economic Complexity Index (ECI) (Ref. 4) which is based on an iterative linear algorithm similar to the Google Page Rank applied to the basket of the exported products of all countries. For the year 2015 the ECI ranks China 38[th] and Qatar 28[th] (Ref. 6), a result which appears totally unrealistic. Qatar exports only Oil and a few related products with minimal diversification, while China is clearly among the few most competitive industrial countries of the world, with an extremely diversified basket of exports. So ECI seems to miss completely the essential elements of growth and industrial competitiveness. This and other similar results (Ref. 5) led to a substantial skepticism towards the whole approach of Economic Complexity. Most of the criticisms up to now have been focused on the limited data used (only exports) and many believe that more extended data should be used to get realistic results. However, we have seen before that the standard analysis of China has been quite incorrect even for institutions which have used many more data.

Our view is that the limited set of data is not a crucial problem, as we will discuss later in more detail. The problem is instead that the ECI algorithm is simply fundamentally inappropriate to capture the desired information from the given dataset. One of the problems is that ECI does not take into any consideration the important role of diversification (Ref. 5), even though this was proposed as one of the motivations for the introduction of ECI (Ref.4): a situation of total confusion. For instance, the authors of (Ref.1) claim that there are various indicators, like ECI and Fitness, which are somehow similar because they are based on a weighted diversification with minor differences in the weights assigned to the various products. This is actually false: in the first case the ECI is just an average and it completely ignores any role of diversification. This is one of the main reasons that motivated us to introduce the new Fitness algorithm (Ref. 2,3), which is precisely the complexity weighted diversification, where the sum of the weights is extensive in the number of products. The Fitness permits to clarify and overcome the fundamental problems of ECI and sets a sound framework for Economic Complexity. Furthermore, we pay special care to implement scientific tests for all our results that provide a clear benchmark to test the various algorithms. In this note we discuss these problems with special attention to the role of diversification.

# 2. Big Data and Economic Complexity

A crucial element of our methodology is a radically new approach to the problem of Big Data (Ref. 12). Big Data is often associated with a subjective ambiguity related to how to consider simultaneously many variables with different units, a procedure that usually leads to the introduction of many arbitrary parameters. In the case of the evaluation of the industrial competitiveness of a country, for instance, the required parameters for such an analysis could be more than one hundred. A key feature of the method of Economic Fitness is to go, in a sense, from 100 parameters to zero parameters and so obtain results which can be tested scientifically. This is done first by focusing on the global structure of the datasets and looking for data for which the signal to noise ratio is optimal and then by developing iterative algorithms in the spirit, but other than Google and, most importantly, tailored for the specific economic aspect one wants to investigate. It is quite naive, indeed, to hope for an all-weather algorithm. In our case, the study of a country or a company is not done at the individual level, but through the global network to which it belongs, introducing a specific non-linear classification algorithm optimized for the global structure of the data, e.g. the triangularity of the country matrix product in the most studied case. In this way, one can obtain the Fitness of the countries and the Complexity of the products, together with many other novel information and strategic insight (Ref. 2,3). Two recent reviews discuss the algorithms for nested bipartite network in some detail (Ref. 13, 14).

The time-dependent information is finally represented in terms of flows in the two-dimensional GDPpc - Fitness plane and analyzed as the evolution of a dynamical system in physical phenomena. As well established in dynamical system theory, for this kind of prediction analysis keeping the dimensionality of the system as small as possible is crucial (Ref. 7,12); consequently, the limited set of data is a strategic choice. Actually, additional data may also be potentially useful, but they should be structured in a hierarchical way, avoiding the construction of too high dimensional spaces of data which usually induce the introduction of too many subjective elements. Since the analysis is made looking at the economy as a global network, an essential mathematical requirement is that the data are homogeneous for the countries and the products. Indeed, this was imposed by the customs and gave rise to the UN-COMTRADE database (Ref. 2,3).

3. **The basic Economic Fitness and Complexity (EFC) algorithm**

Considering the data given by the basket of products exported by each country, the problem is to extract the information on the industrial competitiveness (that we call Fitness) of each country and the Complexity of the products. The starting point is the COMTRADE data organized as a countries-products bipartite network, in which the two types of nodes are joined by undirected links meaning that a given country is competitive on a certain product in the given year. This network is characterized by a pronounced *nestedness* or triangularity (i.e. developed countries are highly diversified, producing all kinds of products including the most complex and less ubiquitous ones, while most poor countries produce only few simple products, those that are exported by all countries, see Fig.1). Nestedness is a general property of ecological, social and economic networks and recently two reviews (Ref. 13,14) give a very useful survey of the field. The Fitness algorithm is discussed in detail as the most appropriate for this class of systems in economics and ecological systems.

Starting with the observation that diversification is an essential element of this network, one can write the first equation in which the Fitness is given by the sum of all the products weighted by their Complexity with no normalization of weights, precisely the concept "rediscovered" in Ref. 1. For the Complexity of the products one can consider that the global nested structure of the bipartite network suggests that in general a high-level product is produced only by few, highly competitive countries. Expressing this concept in mathematics leads to the second equation which is strongly non-linear (bottom panel of Fig.1). For a detailed discussion of this algorithm and the comparison to other ones see Refs. 2, 3 and 5.

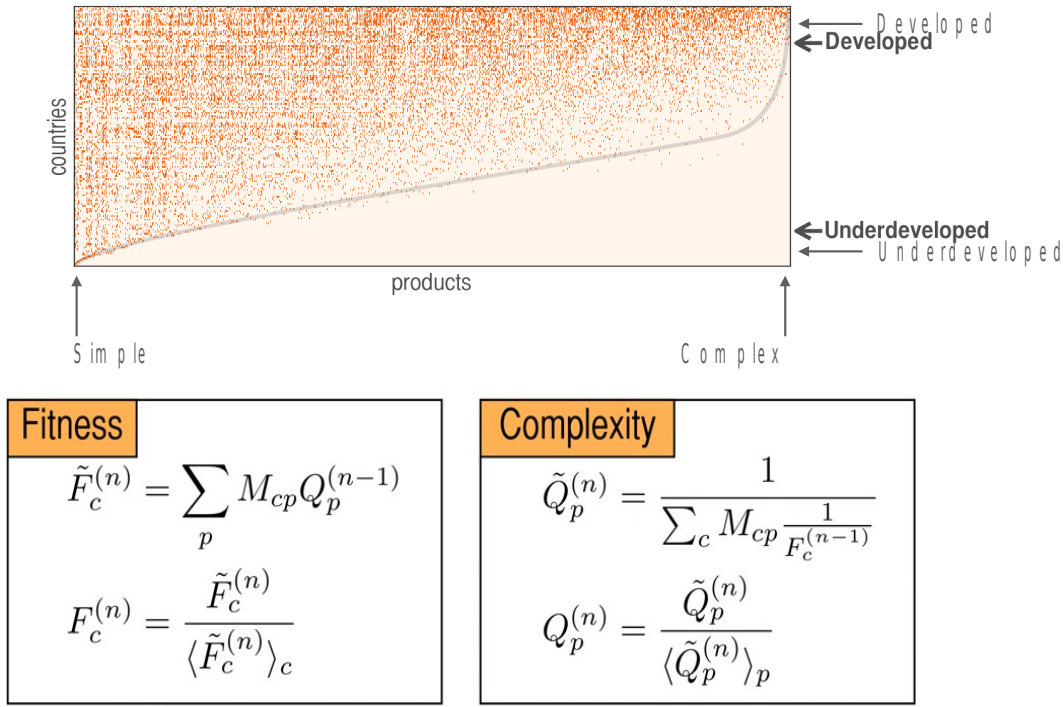

**Fig.1**: Representation of the bipartite network of countries and products. In the vertical axis countries are ranked according to their Fitness (high Fitness at the top) while in the horizontal axis products are ranked according to their complexity (high Complexity on the right). One can see that the countries that produce top products also produce intermediate and low-level products. This implies great importance of diversification and triangular (nested) structure of the data which is also typical of ecosystems. Bottom: The Fitness and Complexity iterative algorithm has been shown to be optimal for nested structures of the type shown here (Ref. 2,3,5,13,14). It is different from the Google Page Rank and it suggests that for each Big Data problem one should identify the appropriate algorithm which better extracts the optimal information.

Let us now consider the situation for the ECI algorithm (Ref. 4). We discuss the ECI in our language to facilitate the comparison with the Fitness. The original presentation of ECI is based on a cumbersome iteration procedure called "The Method of Reflections" which we believe is one of the reasons for the confusion. It can be shown that the ECI and PCI vectors are the solution of a simple linear set of equations (Ref.3):

$$ECI_c^{(N)} = a^{(N)} \cdot \langle PCI_p \rangle_c$$

Equivalent writing of the ECI algorithm: The ECI of a country is proportional to the average PCI of its products,

$$PCI_p^{(N)} = b^{(N)} \cdot \langle ECI_c \rangle_p$$

and the PCI of a product is proportional to the average ECI of the countries that export it.

where the symbols $\langle ... \rangle_x$ indicate the average with respect to x and $ECI^{(N)}$ and $PCI^{(N)}$ indicate the $N^{th}$ eigenvector of the ECI or PCI matrix respectively. For the *a* and *b* coefficients the following relation holds:

$$a^{(N)} b^{(N)} = \frac{1}{\lambda^{(N)}}$$

where $\lambda^{(N)}$ is the $N^{th}$ eigenvalue of the ECI or PCI matrix[1]. It is clear then that there are many equivalent solutions to this problem (all the eigenvectors of the ECI and PCI matrices), and when put in this perspective it is not clear why the case N=2, namely the ECI definition commonly used in the literature and introduced in Ref.4, should be special.

We choose to rewrite the ECI equations in this representation as it is especially clear in showing the absence of any role of diversification. In fact, it is evident that the ECI would be simply proportional to the average PCI of the products, i.e. a sum where the weights are normalized in the export basket of each country. This means that the total number of products is totally irrelevant. If a country produces one thousand products and another country only one with a PCI equal to the average of the thousand products of the other country, their ECI would be the same. It is possible that this misunderstanding about the absence of diversification in ECI arises from the fact that, in the original formulation (Ref. 4), the starting point of the algorithm is indeed diversification but, as soon as one makes the first iteration, the average is taken and any diversification disappears. Moreover, there is no diversification in the second eigenvector technique which corresponds to the fixed point of the algorithm (Ref. 3,5).

Equally important are the problems arising from the second equation because, for example, the Complexity of Oil in the ECI context gets artificially enhanced by the fact that some countries with high Fitness (i.e. the US, the UK etc) are also Oil exporters. As a direct consequence, Oil and other raw materials get wrongly pushed up to relatively high values of Complexity that certainly do not compete to them (Ref.5).

Both these fundamental problems are naturally resolved by the Fitness algorithm. The first equation is based precisely on "Complexity-weighted diversification" and the second equation is such that, as soon as at least one low-level country exports a certain product, this is unavoidably of low Complexity even if it is produced also by some high Fitness countries. This non-linear property is obviously a general feature of economics and, in general, of nested structures and this leads to the noticeable conclusion that economy algorithms should be very different from Google Page Rank. This suggests also that, for each Big Data problem, one should invent the appropriate algorithm taking inspiration by the global structure of data. Consequently, the Big Data related problems are more interesting than just data collection and should include also creative thinking. A recent example is the introduction of the PopRank algorithm to evaluate the impact of Facebook pages and the engagement of the commenting users (Ref. 15).

We are now in the position to understand in detail the differences between the two methods. As we have discussed, the linear relation between product complexity and country Fitness of ECI leads to an unrealistic enhancement of Oil and raw materials. Moreover, ECI considers only the average Complexity of the products and completely ignores, in this way, the very high diversification of China.

For some countries the ECI results may appear somewhat reasonable, but this is due to the fact that, in the real data, there is a correlation between diversification and the average complexity of the products. However, the fundamental lack of any role of diversification in the ECI algorithm becomes evident if one considers the possible suggestions to improve the industrial competitiveness of a given country.

Consider, for instance, a country which makes many products, like China. If one would eliminate all the products except one whose Complexity is above the average (and this is the case for 50% of the products) then its ECI value would increase. So, for example, according to the ECI approach if China would close all its industries except the production of Coalfish, this would enormously increase its ECI value (Fig. 3).

---

[1] Notice that in the typical case where the PCI matrix has a higher dimension than the ECI matrix (and therefore a larger number of eigenvalues), still the number of non-zero eigenvalues is not larger than the dimensionality of the ECI matrix. All the zero eigenvalues correspond to the trivial solution ECI=PCI=0.

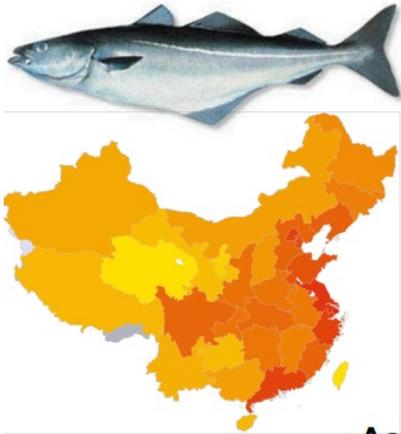

**Fig. 2**: The conceptual inconsistency of ECI to describe industrial competitiveness is made very clear if one considers an example of industrial policy planning based on ECI. The paradoxical result shows that, if China would destroy all its industries except one product, even if not too complex like Coalfish, this would imply an important improvement of ECI. On the contrary, from the perspective of the Fitness, this decision would be disastrous as it should. We dare to hope that this extreme example shows in a definitive way that ECI is totally inappropriate to describe industrial economics and that the two methods are fundamentally different. This example gives also an eloquent answer to the incredible claim that all the algorithms are somehow very similar (Ref.16).

This example shows very clearly the origin of the problems in the ECI and how these are solved by the Fitness. Another problem in comparing data with the official ECI results is that the two original authors of ECI have now separate websites (https://atlas.media.mit.edu and http://atlas.cid.harvard.edu). Looking at the two websites (Feb. 2019) leads to other puzzling observations. In principle they should use the same data and the same ECI algorithm, but for 2016 in the MIT website Saudi Arabia ranks 29th and China 33rd while in the Harvard Blog website Saudi Arabia is 50th and China 18th. Neither one gives any clarification for this macroscopic discrepancy between them. We believe that in this situation it is very difficult to make any substantial progress and for this reason we take special care that our results should avoid any arbitrary parameters, be reproducible, and subject to all possible validity and consistency tests.

### 4. Dynamics and Forecasting

The dynamics in the new GDPpc - Fitness space opens up to a completely new way for monitoring and forecasting (Ref.7). The trajectories of countries in this new space show regions of laminar flow and others characterized by turbulent behavior, which lead to a heterogeneous, non-linear approach to forecasting (Fig.2). This is similar to physical dynamical systems and modern weather forecasting (Ref. 7). In addition to giving a visual insight into the country dynamics, this flow representation has deep implications for the analysis and forecasting. The first conclusion is that standard regressions appear quite inadequate because of the heterogeneous nature of the dynamics. Furthermore, one can adopt the methodology of dynamical systems and use the so-called Method of Analogues (Ref. 12). With this methodology that we call Selective Predictability Scheme (SPS) it has been possible to make forecasts which are competitive and slightly better than those of the IMF (Ref. 8-10).

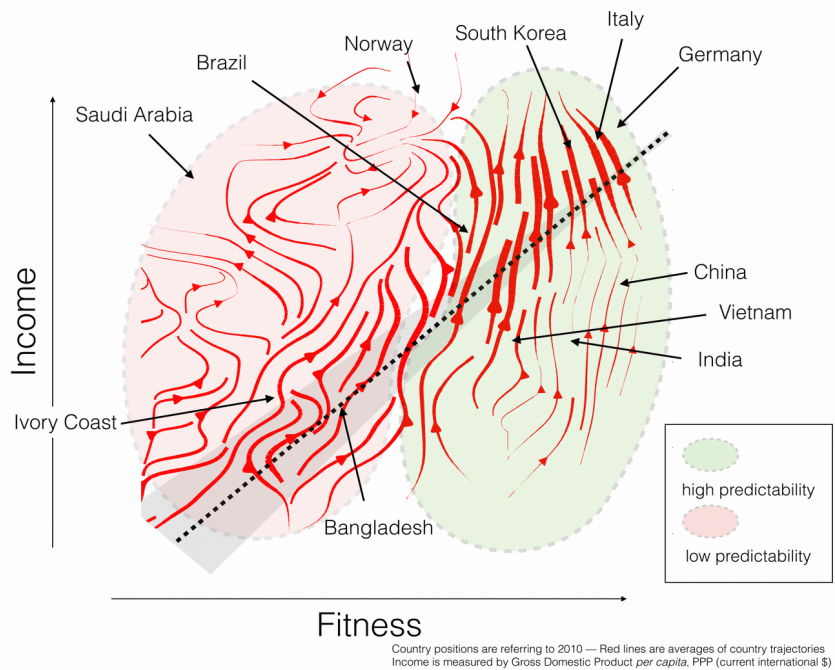

**Fig. 3:** The two-dimensional plane defined by the GDP (per capita) and the Fitness reveals an extremely interesting dynamics for the trajectories of countries. One can observe an interesting heterogeneous behavior characterized by a laminar zone (green) and a turbulent or noisy one (red). This implies that standard regression methods cannot be applied and should be replaced by a new approach strongly inspired by the physics of dynamical systems. This also clarifies the need for dimensional reduction of the problem in relation to the predictability analysis originally formulated by Poincaré and recently studied by various authors (Ref.12).

In view of this, it is clear that the authors of (Ref.1) borrowed also this two-dimensional dynamical approach from our papers. In this paper ECI is replaced by a Principle Component Analysis (PCA) which goes in the right direction of a complexity weighted diversification, but the results are not clear. Indeed, a ranking of the countries in terms of PCA is not even reported and even the authors conclude that their data can only give a qualitative insight. On the contrary, our methodology has been extensively used for detailed forecasting. This situation is totally misrepresented in the references of Ref.1.

5. **Statistical validation of scientific results.**

Since our strategic objective is a radically novel approach to fundamental economics in the direction of a more scientific discipline, the problems of validation and forecasting represent an essential element. This implies avoiding as much as possible arbitrary parameters in the whole analysis which is then scientific and reproducible. This results in a clear methodology for the validation of the results and the associated forecasting power. The forecasting potential has been already discussed for its intrinsic value and importance and it is the object of a recent paper (Refs.8 and 10). Controlled and out-of-sample forecasting represents a very important "experimental" test for the validation of the whole methodology which we will apply in detail to all our results.

On the contrary the results of Hidalgo and Hausmann groups were validated only by some sort of regressions which lead to intrinsically inconsistent results. For example, about two years ago, in the first assault to the Fitness algorithm, Hidalgo and collaborators claimed to have finally identified the best algorithm for Economic Complexity and they called it "ECI+" (Ref.17). In fact, their regressions showed that ECI+ performed better than ECI which, on its turn, was better than the Fitness. After a couple of weeks, we pointed out that ECI+ is just a renaming of the Fitness and the two are actually mathematically the same algorithm (Ref.18). Their reaction was a new note in which they argued that, after all, the various algorithms are all alike (Ref. 16). So, their regressions can show that the same algorithm works well if it is called ECI+, but not if it is called Fitness. In addition, after a few weeks, the conclusion is totally different and all algorithms become equal if compared by the same kind of regressions (Ref.4). How is it possible that these regressions give different results for the same algorithm with different names and again completely different after two weeks was never clarified. This led to a debate in which four papers appeared in ArXiv, two from Hidalgo's group and two from our group. In chronological order, these are the Refs. 17,18, 16 and 4. It is not exactly elegant that in the new paper of Hausmann's group (Ref.1, the second assault to the Fitness) out of a discussion which consisted of four papers, only the two papers from Hidalgo's group are mentioned. In this respect, it is interesting to note that after we introduced the Fitness algorithm for some time the reaction of Hausmann, Hidalgo and collaborators has been to ignore and dismiss it completely. Recently, however, the situation has taken a new twist. On one hand, both groups try to hide the fundamental flaws of ECI and the differences between the two algorithms, arguing that the various algorithms are actually very similar. On the other hand, each of them rediscovers the Fitness in various ways and presents it as the new concept which really works. This whole situation certainly does not help to develop a fair scientific discussion and to make real progress in the field.

Finally, a constructive note. Stimulated by the fundamental observation of the fantastic growth of China, we made an effort to understand this phenomenon in detail (Ref.19). In doing this we noticed a sort of complementarity between the Fitness methodology and the New structural Economics developed by Justin Lin and collaborators (Ref. 20,21). The first method provides a scientific, data-driven ground for the economic considerations of the second.